\documentstyle[11pt]{article}
\begin{document}

\def\beq{\begin{equation}}
\def\eeq{\end{equation}}
\def\bce{\begin{center}}
\def\ece{\end{center}}
\def\bea{\begin{eqnarray}}
\def\eea{\end{eqnarray}}
\def\ben{\begin{enumerate}}
\def\een{\end{enumerate}}
\def\ul{\underline}
\def\ni{\noindent}
\def\nn{\nonumber}
\def\bs{\bigskip}
\def\ms{\medskip}
\def\wt{\widetilde}
\def\wh{\widehat}
\def\brr{\begin{array}}
\def\err{\end{array}}
\def\Tr{\mbox{Tr}\ }

\thispagestyle{empty}
\renewcommand{\thefootnote}{\dagger}

\vskip 0.5truecm

\centerline
{\Large \bf Perturbative approach to the two-dimensional}
\vskip 0.3truecm

\centerline
{\Large \bf quantum gravity}

\vskip 1.6truecm

\centerline
{{\large \bf I. L. Shapiro}\footnote{Talk given at the International
Seminar "Quantum Gravity" (Moscow, June 11 - 21, 1995)}}

\vskip 0.4truecm

\centerline{\sl Departamento de Fisica Teorica, Faculdad de Ciencias,
Universidad de Zaragoza,}

\centerline{\sl 50009, Zaragoza, Spain}
\vskip 2mm

\centerline{\sl and}
\vskip 2mm

\centerline{\sl Department of Mathematical Analysis}

\centerline{\sl Tomsk State Pedagogical University, Tomsk, 634041, Russia}

\vskip 1.6truecm

\centerline{\large \sl Abstract}
\vskip 0.4truecm
\noindent
The main part of this presentation is a review of the
original works \cite{23,29,ShB,modphys,npsi,anom} on the perturbative
approach to the $2$-dimensional quantum gravity.
We discuss the renormalization of the
two-dimensional dilaton gravity in a covariant gauge, the form of the
quantum corrections to the classical potential, and the conditions of
Weyl invariance in a theory of string coupled to $2d$ quantum gravity.

\setcounter{page}1
\renewcommand{\thefootnote}{\arabic{footnote}}
\setcounter{footnote}0
\vskip 20mm


\noindent
{\large \bf 1. Introduction}
\vskip 5mm

Recent progress in a nonperturbative formulations of quantum
field theory leads to the increasing interest in this field.  However
at the moment most of the achievements concerns some special simple
models which are essentially different from the ones which can be applied to
describe the phenomenology. On the countrary any phenomenological models are
rather complicated and therefore it is not clear whether it is possible
to obtain any rigour nonperturbative information about them.
Thus the standard perturbative way of study is still relevant because in
a lot of cases it provides us by the results. In such a situation it is
quite natural to apply the standard perturbative methods to the more simple
models and thus to get deeper understanding of their internal
distinguished properties.

For example here we present some results of the perturbative study of
the two-dimensional quantum dilaton gravity. Some versions of the
theory are linked with the effective induced action of string on the
curved wourld sheet \cite{1}, they are
exactly soluble \cite{15,16} and therefore it is natural
to address the following questions:

i) Is it possible to distinquish these versions perturbatively?

ii) Are there some other models with the similar properties?

iii) What is the peculiarity of the two-dimensional gravity which provides
it's simple nature?

iv) Since the two-dimensional gravity is closely related to the theory of
the noncritical string, is it possible to obtain some extra information
about the last in the framework of such a study?

Here we shall try to answer these questions.
In fact some of them look a bit naive, especially the third
one. It is well known that the two-dimensional metric has only one
(conformal, for instance) degree of freedom, and it is just the main
feature of the two-dimensional
gravity. However the problem is not so simple
because the standard perturbative approach supposes the general covariance
to be preserved at quantum level. In a covariant gauge the
two-dimensional metric has three degrees of freedom that is compensated
by the FP ghosts. Thus it is very interesting to understand how the
special feature of the two-dimensional gravity looks in this gauge.
As it will be shown below the main property of the two-dimensional
gravity manifest itself in the form of some explicit identity. This
identity concerns the components of covariant quantum metric and leads
to some properties related with the especially strong
dependence on the gauge fixing parameters.

The starting point of our discussion is the theory of
$D$ copies of the
free massless scalar fields coupled to the $2$-dimensional metric.
As usual, we suppose that the integration over the string coordinates
and the reparametrization ghosts is performed before the integration over
the conformal factor \footnote{Another order of integrations will be
discussed in the consequent sections.}.
Vacuum guantum effects and the contribution of the
reparametrization ghosts lead to the conformal anomaly and the VEV of the
trace of the energy-momentum tensor becomes \cite{Duff}.
\beq
T=<T_{\mu}^{\;\;\mu}>=\frac{1}{16\pi}\;aR,\;\;\;\;\;\;\;\;\;\;\;\;
a = \frac{D-26}{6}    \label{1x}
\eeq
The anomaly which appears in the noncritical
dimension of the target space, leads to the equation for the effective action
\beq
-{2\over\sqrt{-g}}\; g_{\mu\nu}\;{\delta\Gamma\over{\delta g_{\mu\nu}}}=T
\label{2x}
\eeq
which can be solved in the following nonlocal form \cite{1}.
\beq
 \Gamma[g_{\mu\nu}] = \frac{1}{16\pi}\;a\;
\int d^{2}x \sqrt{-g_{x}} \;\int d^{2}y\sqrt{-g_{y}}\;
R(x)\;\left( {\frac{1}{\Box}}\right)_{x,y}\; R(y)      \label{3x}
\eeq
The last action can be rewritten in the local form with the help of
the dimensionless auxiliary scalar $\Phi$.
\beq
S_{g} = \int {d^2}\sigma \sqrt {g}\; \left\{ {1\over 2}\; g^{\mu\nu}
\partial_{\mu}\Phi\partial_{\nu}\Phi + C_{1}R\Phi + V(\Phi)\right\}\label{4x}
\eeq
with
\beq
C_1= \frac12\;\left( \frac{a-1/6}{2\pi} \right)^{1/2},
\;\;\;\;\;\;\;\;\;V(\Phi)=0  \label{5x}
\eeq
Such an action describes the induced $2d$ gravity for the noncritical
string theory, and therefore one must take into account it's quantum effects.
Indeed one can start with the generalized theory and regard $ C_{1}$ as an
arbitrary constant and $V(\Phi)$ as an arbitrary function
of the field $\Phi$. This is just that we shall do. Below we
discuss in details the renormalization of the theory (\ref{4x}) in an
arbitrary covariant gauge as well as in the conformal gauge. Then we
consider the effective potential and establish the gauge dependence
of quantum corrections. The methods of calculations developed here turns
out to be very useful and enables one to consider the integration over the
string coordinates and the two-dimensional metric without special order.
In this case the action of the sigma model in a background fields
can be regarded as the direct generalization of (\ref{4x}). The use of
covariant gauge enables us to explore the role of the special
order of integrations in a consistent manner.

There were a lot of interesting and important investigations in the
field of $2$ and $2+\varepsilon$ dimensional quantum gravity
(see, for example, \cite{12a,12b,12,13,14,15,16,13,14,Cha,21,KKN}).
The perturbative approach has been developed in
\cite{22,23,29,ShB,modphys,npsi,anom,24,24a,24b}.
Here we review and discuss the results of
\cite{23,29,ShB,modphys,npsi,anom}.

\vskip 8mm
\noindent
{\large \bf 2. The action of induced gravity and it's transformation
properties}
\vskip 3mm

As it was pointed out in \cite{23,24} the different forms of the action
for the two - dimensional dilaton quantum gravity are equivalent to
the expression (\ref{4x}).
Since this fact has direct relation to our study let us consider it in
some details. One can introduce another scalar $\varphi$ and also
perform the conformal transformation of the metric
\beq
g_{\mu\nu}
={\bar g}_{\mu\nu}\;e^{2\sigma(\varphi)}             \label{6x}
\eeq
In terms of new variables the starting action (\ref{4x}) becomes
\beq
S_n = \int {d^2}\sigma \sqrt {{\bar g}}\; \left\{ {1\over 2}
{\bar g}^{\mu\nu}\left[\Phi'^2+4C_1\Phi'\sigma'
\right]
\partial_{\mu}\Phi \partial_{\nu}\Phi + C_{1}{\bar R}\Phi + e^{2\sigma}
V(\Phi)\right\} \label{7x}
\eeq
where $\Phi=\Phi(\varphi),\;\sigma = \sigma(\varphi)$ and the primes denote
the derivatives with respect to $\varphi$. Thus it is clear that
the actions
\beq
S = \int {d^2}\sigma \sqrt {g}\; \left\{ {1\over 2} g^{\mu\nu} g(\Phi)
\partial_{\mu}\Phi \partial_{\nu}\Phi + c(\Phi)R + U(\Phi)\right\}
 \label{8x}
\eeq
with an arbitrary functions $g(\Phi)$ and $c(\Phi)$
are linked by the conformal transformation of the metric supplemented
by the reparametrization of the dilaton,
and the corrresponding change of the
potential function. In particular one can consider the theory
\beq
S_{g'}=\int {d^2}\sigma\sqrt {g}\;\left\{C_{1}R\phi+v(\phi)\right\}\label{9x}
\eeq
which is classically equivalent to the original theory (\ref{4x}) (with
accuracy to some change of the potential, that will be established below).

One can easily see that the above transformation has direct relation to the
renormalization of the theory. If we work in the covariant formalism,
then all the possible divergences are local covariant expressions. Taking into
account the power counting it is easy to prove that the possible
divergences in the theory (\ref{4x}) have the structure similar to (\ref{8x}).
\beq
\Gamma_{div} = \int {d^2}\sigma \sqrt {g}\; \left\{ {1\over 2} g^{\mu\nu}
A_1(\Phi)\partial_{\mu}\Phi\partial_{\nu}\Phi+A_2(\Phi)R +
A_3(\Phi)\right\}                  \label{10x}
\eeq
Since (\ref{4x}) and (\ref{8x}) are conformally equivalent, the first one is
renormalizable. Moreover one can establish the renormalzation of the
the "more general" theory (\ref{8x}) with the help of divergences calculated
in a "particular case" (\ref{4x}). For instance, one can start with (\ref{8x}),
transform it to (\ref{4x}) with different field variables and than
perform an inverse transformation of the variables in an expression
for the divergences. Indeed this inverse
transformation can be affected by quantum corrections.
However at one loop order it is possible to use the
naive classical invertion.
\footnote{We remark that it is not obviously true for the dimensions higher
than two.}

\vskip 8mm
\noindent
{\large \bf 3. The one-loop calculations}
\vskip 3mm

To perform the one-loop calculations we apply
the background field method and the standard Schwinger-DeWitt technique
of extracting the divergences. The special features of the theory
requires the special choice of the gauge fixing
condition which enables us to investigate the spectrum of the
differential operators of the nonstandard form.

Let's make the
splitting of the fields $g_{\mu\nu}$ and $\Phi$ into the background
and quantum parts.
$$
 g_{\mu\nu}\;{\rightarrow}\;{g'_{\mu\nu}} =
g_{\mu\nu} + h_{\mu\nu}, \;\;\;\;  h_{\mu\nu} = {\bar h}_{\mu\nu} +
{1\over 2}hg_{\mu\nu},
\;\;\;{\bar h}_{\mu\nu}{g^{\mu\nu}} = 0
$$
\beq
\Phi \;{\rightarrow}\;\Phi' =\Phi + \phi
\label{11x}
\eeq
Here $g_{\mu\nu}$ and $\Phi$ are background fields,
the ${\bar h}_{\mu\nu}$ is the traceless part of the quantum metric.
We divide the quantum metric into trace $h$ and traceless part for the
sake of convenience.
The one - loop effective action is defined by the bilinear (with
respect to the quantum fields) part of the classical action.
Since the corresponding bilinear form is degenerate,
we have to introduce the gauge fixing term and the action of ghosts.
The more general form of the covariant gauge fixing term is following:
\beq
 S_{gf} = - \int {d^2}\sigma \sqrt
{g}\;{\chi}_{\mu}\; {G^{\mu\nu}}\;{\chi}_{\nu} \label{12x}
\eeq
where
${\chi}_{\mu}$ is the background gauge and $G^{\mu\nu}$ is the weight
operator.
\beq
{\chi}_{\mu} = {\nabla}_{\nu}{{\-h}_{\mu}^{\nu}} -
{\beta} {\nabla}_{\mu} h - {\gamma}{\nabla}_{\mu}\phi -
E_{\mu}^{\rho\sigma} h_{\rho\sigma} - F_{{\mu}}\phi,
\;\;\;\;\;\;\;
G^{\mu\nu} = {\tau} {g^{\mu\nu}} \label{13x}
\eeq
Here ${\tau},\; {\beta},\; {\gamma}, \;E_{\mu}^{\rho\sigma},\;
F_{\mu}$ are arbitrary functions (gauge parameters)
of the dimensionless background field $\Phi$.
Let us now make some remark conserning the choice of these
functions. For the sake of simplicity we require the bilinear
form of the total action $S^{(2)} + S_{gf}$ to be minimal.
The last means that the second derivatives of the quantum fields appear
only in the combination $\Box =\nabla_\nu \nabla^\nu$. In the four
dimensional gravity such a condition fixes  all the gauge parameters
(see, for example, \cite{38}). On the
countrary, in the case of the two-dimensional gravity we can apply the
identity proved in \cite{anom} (see also another proof in \cite{ShB})
for the particular case of the flat background metric)
\beq
{\bar h}^{\mu\nu}\;X\;[\;\frac{1}{2}\delta_{\mu\nu,\alpha\beta}\Box -
g_{\nu\beta}
\nabla_{\mu}\nabla_{\alpha} - \frac{1}{2}\delta_{\mu\nu,\alpha\beta} R\;]\;
 {\bar h}^{\alpha\beta} = 0                              \label{14x}
\eeq
Here $X$ is an arbitrary scalar function.
It turns out that (\ref{14x}) allows us to get the minimal
operator for an arbitrary value of some gauge parameter $\nu$.
At the same time the gauge
 parameters ${\tau},\; {\beta}$ and ${{\gamma}_a}$ must be taken in a
special way to provide the minimality of the total action.
\beq
 {\beta} = 0, \;\;\;\;\;
{\gamma} = -{\nu \over{\Phi}},\;\;\;\;\;
{\tau} = {{\Phi}\over{\nu}}                                 \label{15x}
\eeq
Furthermore there remains an
arbitrariness related with the functions $E_{\mu}^{\rho\sigma}$ and
$F_{\mu}$. The explicit calculation shows that the divergent part
of the one - loop effective action does not depend on these functions
\cite{23}. In fact the lack of such dependence can be proved in a general form,
that will be demonstrated below.

For the special "minimal" choice of the gauge fixing the bilinear part of
the total action $S + S_{gf}$ has the form
\beq
(S + S_{gf})^{(2)} = \frac{1}{2}\int {d^2}\sigma \sqrt {g}\;({\bar
h}_{\mu\nu},\;h,\;\;\phi ) (\hat{H})({\bar
h}_{\alpha\beta},\;h,\;\;\phi)^T                            \label{16x}
\eeq
where $T$ denotes
transposition, and the self - adjoint operator ${\hat H}$ has the structure
\beq
{\hat H} = {\hat K} {\Box} + {\hat L}^{\lambda} {\cal D}_{\lambda} +
{\hat M}                                                      \label{17x}
\eeq
Here ${\hat K}, {\hat L}^{\lambda},{\hat M}$
are c-number operators acting in the space of the fields
$({\bar h}_{\rho\sigma},\;h,\; \phi)$
and $det||{\hat K}||$ is nonzero. One can find an explicit expressions in
\cite{29}.

The action of the Faddeev-Popov ghosts is defined in a usual way and
has the form \cite{29}.
\beq
 S_{gh} = \int
{d^2}\sigma \sqrt {g}\;{\bar C}^{\alpha}\;M_\alpha^\beta\;C_\beta \label{17ax}
\eeq

The one - loop divergencies of the effective action are given by
the expression
\beq
\Gamma^{(1)}_{div} = -{1\over 2} \left.Trln{\hat H} \right|_{div} +
 \left.Trln{\hat M}_{gh} \right|_{div}                           \label{18x}
\eeq

The form of the operator ${\hat M}_{gh}$ enables us to apply
the standard Schwinger-DeWitt technique of the divergences calculation.
On the other hand,
the structure of the operator ${\hat H}$ (\ref{17x}) enables us
perform the following transformation.
\beq
Trln{\hat H} = Trln{\hat K} + Trln ({\hat 1} \Box + {\hat
K}^{-1}{\hat L}^\lambda
\nabla_\lambda + {\hat K}^{-1}{\hat M})              \label{19x}
\eeq
First term in (\ref{19x}) does not give contribution to the
divergences, because ${\hat K}$ is local operator and therefore this
term can be
omitted.  The second term  has  standard structure as well as
the ghost action operator.
Now it is possible to derive the one-loop divergences, applying the well-known
result of the Schwinger-De Witt expansion. The resulting values of
$A_1, A_2, A_3$ are \cite{29,anom}.
\beq
A_{1} = - {\nu\over{{\varepsilon}{\Phi}^2}},\;\;\;\;\;\;\;\;\;\;
A_{2} = {1\over{6\varepsilon}}\;R,\;\;\;\;\;\;\;\;\;\;
A_{3} = {1\over{\varepsilon}} \left[ {{\nu}V\over{C_{1}{\Phi}}} +
V'{1\over {C_{1}}}\right]                       \label{20x}
\eeq
where $e=2{\pi}(d-2)$ is the parameter of dimensional regularization.
Let us make some comments on the above result.

i) The gauge dependence of the divergences is proportional to the
dynamical equation $g_{\mu\nu}\; \frac{\delta S_g}{\delta g_{\mu\nu}}$.
Therefore it can be removed by the renormalization of the metric.
The remaining divergences are in the potential $A_3$ sector and
also the topological Einstein divergences in $A_2$ sector.
On shell the one-loop divergences are gauge independent as it has to be.

ii) From the previous point it follows that the divergences in the kinetic
sector are essentially determined by the ones in the potential sector.
Since the gauge fixing parameters $E,F$ (\ref{13x}) may affect only
the kinetic type divergences and therefore any dependence on these
parameters is forbidden. We note that this follows from general backgrounds
and also  agrees with the result of direct calculations.

iii) In the conformal gauge the theory (\ref{4x}) becomes the linear $D=2$
sigma model with the nontrivial tachyon sector only \cite{24}.
It is well known that in such a model only the tachyon type divergences
may arise, and therefore the above result can be regarded as the confirmation
of this known fact. Indeed we have used the covariant gauge where (\ref{4x})
is not equivalent to the ordinary sigma model. In this sence our calculation
gives more general information.

iv) The coefficient of the pure gravitational diveregence $A_2$
also corresponds to the $D=2$ sigma model. It is remarcable that it
can be calculated in a direct way in a harmonic gauge,
without the study of the reparametrization ghosts $etc$ as it have been done
in \cite{1}.
The calculation of this counterterm is
essentially based on the identity (\ref{14x}) and shows that this coefficient
does not depend on the gauge parameter $\nu$.
One can see that the one-loop divergences of the Einstein type are
uniquely defined in covariant gauge. Thus one can consider the renormalization
of the composite operator $<T_\mu^{\mu}>$
 and obtain the expression
for the conformal anomaly in a covariant gauge.
We remark that
the $R$-type counterterm is defined in a unique way, and therefore
the corresponding trace anomaly is defined in a unique way in the
harmonic gauge. In this sence our result exactly corresponds to the
calculation of the anomaly performed earlier in a harmonic gauge in
\cite{di,flp,rk,kr,bb} and also discussed in \cite{mo}. Indeed the calculation
of the counterterm is essentially more simple if we take into account
the identity (\ref{14x}).
Below a more general calculation of $<T_\mu^{\mu}>$ will be
considered in some details.

At the end of this section we discuss the
one-loop divergences of the more general dilaton theory
 (\ref{8x}). According to our previous consideration
one can extract these divergences transforming (\ref{20x}).
It is possible to
verify this fact with the help of some simple particular case.
To do this let us consider the conformal transformation between the
 model (\ref{4x}) and the model
without kinetik term for the dilaton (\ref{9x}).
Direct calculation
performed in the model (\ref{9x})
shows that in the minimal gauge the divergences has exactly the
same form (\ref{20x}).
This makes the analysis fairly easy.
The conformal transformation, reparametrization of the dilaton and the
transformation of the
potential functions which link (\ref{4x}) and (\ref{9x}) are
[for the sake of convenience we denote the metric of the model (\ref{4x})
as ${\bar g}_{\mu\nu}$, see (\ref{6x})].
$$
\Phi = \phi + \phi_0, \;\;\;\;\;\;\;\; \phi_0 = const.
$$$$
\sigma = \frac{1}{4C_1}\;\phi + \sigma_0,\;\;\;\;\;\;\;\; \sigma_0 = const.
$$
\beq
v(\phi) = V(\Phi) exp \left( \frac{1}{2C_1}\Phi + 2\sigma_0 \right)
\label{21y}
\eeq
It is useful to put $\sigma_0 = \phi_0 = 0$.
The inverse transformation corresponds to the conformal factor $-\sigma$
and to $V=v e^{-2\sigma}$.

Now we have two ways to derive the one-loop divergences in the theory
(\ref{9x}) in a harmonic gauge. The first one is the
direct calculation similar
to that we have just presented for the theory (\ref{4x}). The second
way is
to change variables according to (\ref{21y}), then use
(\ref{20x}) and then
make an inverse transformation. After some calculations
we arrive at the following results. As it was already mentioned,
 on the first
way we obtain (\ref{20x}). The calculation in transformed field variables
gives different result in the potential sector
\beq
{\bar A}_3 = {1\over{\varepsilon}} \left[ {{\nu}V\over{C_{1}{\Phi}}} +
V'{1\over {C_{1}}} - \frac{V}{2{C_1}^2}
\right]                       \label{22y}
\eeq
The origin of this difference is the different choice of the field
quantum variables. One can easily check that  on the classical
equations of motion the difference in two expressions concerns only
the surface counterterm $\Box \Phi$. And so we have seen that our
consideration of the renormalization of the theory (\ref{8x}) was true.
One can establish the corresponding counterterms
using the result (\ref{20x}) computed for the model (\ref{4x})
and the formula (\ref{6x}).

\vskip 8mm
\noindent
{\large \bf 4. The effective potential and gauge dependence}
\vskip 3mm

It is a well know fact that in the gauge
 theories the effective potential
depends on the gauge fixing parameters. In the
two dimensional dilaton gravity such a dependence is even stronger than
in other cases.  We define the effective potential $V_{eff}({\Phi})$
as the part of
effective action which survives on the constant background
${\Phi} = const, R=0$.
So $A_{3}$ in (\ref{20x}) is the divergent part of
$V_{eff}({\Phi})$.
One-loop effective action is given by the expression
\beq
V_{eff}^{(1 - loop)} = V - {\Delta}V - {1\over 2}
Tr \sum_{k=1}^{4}ln{\lambda}_{k} +
Tr \sum_{l=1}^{2}ln{\lambda'}_{l}       \label{21x}
\eeq
where ${\lambda}_{k}$ and ${\lambda'}_{l}$ are the eigenvalues
of the operators ${\hat H}$ and ${\hat M_{gh}}$
respectively. Tr includes the integration over the momentums in the
framework of some regularization scheme.
(\ref{21x}) together with the identity (\ref{14x}) give the possibility
to derive the one-loop effective potential for arbitrary gauge fixing
parameters $\nu, \beta, \gamma$. One can find an explicit most general
expression for $V_{eff}$ in \cite{ShB} where the
cut-off regularization have been used.
Just as in the conformal gauge \cite{24}
the logarithmical divergences of the effective
potential $V_{eff}$ depend on $V''$ as well as the finite part,
but only for the nonminimal gauges.

However the most interesting
is the minimal gauge $\beta = 0,\; \gamma = {\nu}{\Phi}^{-1}$.
As it was already noted above, the identity (\ref{14x}) leads to that
the "natural" configuration space metric in the
theory (\ref{4x}) depends on the gauge parameter $\nu$.

Taking the counterterm in the form
\beq
{\Delta}V = + {1\over4{\pi}} {\mu}^{2} A({\Phi})
ln\left({{\Lambda}^{2}\over{{\mu}_{2}}}\right) + {1\over4{\pi}}
{\Lambda}^{2} B({\Phi}),   \label{22x}
\eeq
where ${\Lambda}$ is the cut-off regularization parameter,
 ${\mu}^{2}$ is the dimensional parameter of renormalisation,
and $A,B$ are some unknown functions, we require $V_{eff}^{(1 - loop)}$
to be finite and after determination of $A$ and $B$ we finally get
\cite{ShB}
\beq
V_{eff}^{(1 - loop)} = V - {1
\over4{\pi}}
\left\{ {V'\over{C_{1}}}\left[1 - ln\left({V'\over{C_{1}{\mu}^{2}}}\right)
\right] +
{{\alpha}V \over{C_{1}{\Phi}}}
\left[ 1 - ln \left( {{\alpha}V \over{C_{1}{\Phi}{\mu}^{2}}}\right)
\right] \right\}                                            \label{23x}
\eeq

     The expression (\ref{23x})
 has to be supplemented by the
normalization conditions. Let us, for example, consider the exactly
soluble case $V({\Phi}) = {\Omega}{\Phi}$. Then the quantum
correction in  (\ref{23x}) is some
 constant which doesn't depend on ${\Phi}$. Introducing the
condition
\beq
V_{eff}^{(1 - loop)}({\Phi} = 0) = 0                   \label{24x}
\eeq
we find $V_{eff}^{(1 - loop)} = V$ that is the absence of all
the quantum corrections.
     For the other interesting case $V_{eff}^{(1 - loop)} =
{\Omega} exp({\lambda}{\Phi})$ we take the
normalization condition of the form
\beq
V_{eff}^{(1 - loop)}({\Phi} = 0) = {\Omega} \label{25xa}
\eeq
and find that quantum corrections do not vanish even for the
value ${\alpha} = 0$, and do not repeat the structure of $V$.

     Since we are interested here in the general form of the
potential function which provides the triviality of the one-loop
corrections, the gauge dependence of this quantity leads to some problem.
Since $V_{eff}$ is gauge dependent it is not clear what means the
correct choice of gauge. Thus it is natural to try to use the unique
effective action of Vilkovisky \cite{19b} which is supposed to give the
gauge-independent result. At the same time there is the well - known
problem in the unique effective action which contains the dependence on
the metric in the space of fields (configuration space) \cite{23b,28b,31b}.
The additional condition of Vilkovisky \cite{19b} fixes this metric to
be natural (that is the matrix in the bilinear form of the action in a
minimal gauge). This condition implies the essential use of the
classical action of the theory. However in the theory (\ref{4x}) even the
natural configuration space metric depends on the gauge parameter
$\;\nu\;$\thefootnote{The problem with a choice of the "natural" metric
in $d=2$ gravity has been pointed out already in the original work of
Vilkoviski \cite{19b}.}
and therefore such dependence may arise in the unique effective potential.

The direct calculations have shown,
that the gauge dependence of the Vilkovisky
effective potential really takes place \cite{ShB}. Note that it is also
follows from the calculations of Kantowski and Marzban \cite{24a} with the
natural metric of the form:
\beq
G_{ij}=\left(\matrix{
{C_{1}{\Phi}\over{2\nu}} &0                &0\cr
 0                          &0                &-C_{1}/2\cr
 0                          &-{C_{1}\over 2}
&(1-{C_{1}{\nu}\over{\Phi}})
\cr}\right)
\label{25x}
\eeq
that corresponds to the matrix ${\hat K}$ in (\ref{17x}).
Thus the gauge dependence of the effective potential in the theory (\ref{4x})
is even stronger than usually, in the sence that it can not be removed
by taking into account the Vilkoviski corrections.
Thus the only one thing we can do is to choose some special "correct"
gauge. In view of the special form of the gauge dependence this corresponds
to some special conformal reparametrization of the background metric.

Let us now turn to the search of the theories with the trivial
quantum corrections to $V({\Phi})$. We shall take the condition of
triviality in the form:
\beq
V_{eff}^{(1 - loop)} = (1 + {\tau})V + {\eta} \label{26x}
\eeq
where ${\tau},{\eta}$ are some constants. If we choose the gauge with
the vanishing kinetic
divergent contributions to the effective action then
 there are only three appropriate gauges: i) light - cone,
ii)conformal and iii)harmonical gauge with ${\nu}$ = 0.
Since in the last case the loop corrections to the
effective potential look more simple let us consider iii). Then
(\ref{26x}) is rewritten in the form
\beq
{\tau}V + {\eta} = {V'\over{4 {\pi} C_{1}}}
ln\left({V'\over{eC_{1}{\mu}^{2}}}\right)                    \label{27x}
\eeq
This is an ordinary differential equation which can
be solved easily. In the case of ${\tau}$ = 1 there is only one
(well - known) solution $V({\Phi}) = {\Omega}{\Phi} + {\Omega}_{1}$.
where ${\Omega},{\Omega}_{1}$ are some constants.
However if ${\tau}$ is not equal to 1 there are two additional solutions of
the form:
\beq
V = - {{\eta} \over {\tau}}
+ {{\mu}^{2}\over{4{\pi}{\tau}}}\;\left[- 1 {\pm} \sqrt {8{\pi}
C_{1}{\tau}({\Phi} - {\Phi}^{*})} \right]\; e^{\pm \sqrt{8{\pi}
C_{1}{\tau}({\Phi} - {\Phi}^{*})}}   \label{28x}
\eeq
Here ${\Phi}^{*}$ is the constant of integration.
The expression (\ref{28x}) contains an arbitrary parameters ${\eta},{\tau}$
and moreover the extra dependence on the renormalization parameter
${\mu}$. The last dependence has to be removed by the introduction of
some normalization conditions. If one uses the condition (\ref{26x})
then the nontrivial part of the renormalized effective potential
turns out to be zero.
On the other hand it is much more natural to take the more
general normalization condition of the form
\beq
V_{eff}^{(1 - loop)}(0) =
 {\tau}^{*}V(0) + {\eta}^{*}   \label{29x}
\eeq
where ${\tau}^{*}, {\eta}^{*}$ are some constants (which may
coincide with ${\tau}, {\eta}$). Then it is possible to solve (\ref{29x})
and to express ${\mu}^{2}$ in terms of ${\tau}^{*}, {\eta}^{*},
{\tau}, {\eta}$.

\vskip 8mm
\noindent
{\large \bf 5. Covariant calculation for string coupled to quantum gravity}
\vskip 3mm

In standard string theory it is accepted to consider the path integral
in two steps \cite{1}. First the integration over string coordinates is
fulfilled and then the one over the two-dimensional
metric $g_{\mu\nu}$. After the
first integration the condition of Weyl invariance is introduced and
after that the intergal over the two-dimensional metric is reduced
to the integral over only modulus.
The conditions of Weyl invariance produce the effective equations
for the background fields, these equations appears as a series in $\alpha'$
\cite{2,call,3,tse,osb,37}.
In what follows we shall name this approach as the standard one.

It is usually  assumed that the result does not depend on the order of
integrations (see, for example, the papers \cite{31,32}).
In both papers the conformal gauge has been used.
In this gauge the $D$ dimensional sigma model coupled to
quantum gravity can be transformed to the $D+1$ dimensional
sigma model on classical
gravitation background whereas the conformal factor of the metric plays
the role of an extra coordinate. The qualitative analysis of the
quantum gravity effects leads to conclusion that the Weyl invariance of the
theory puts additional restrictions on the starting  $D$ dimensional
sigma model \cite{31}. More detailed analysis including the qualitative
account of contribution of the Jacobian of the conformal transformation
shows that all the effects of quantum gravity concerns the additional
 $D+1$ dimensional couplings which have to be introduced for the sake of
 renormalizability. As a result the effective equations in original $D$
dimensions are the same as in the case when gravity is classical
background \cite{32}.

All the conclusions of \cite{31,32}
are based on the use of conformal gauge. In this  gauge the
renormalizability may be broken by the terms, depending on conformal factor.
Therefore it is quite
reasonable to explore the problem in a covariant gauge where the
renormalization structure is fairly simple \cite{anom}.
Our purpose is to investigate whether the quantum
gravity leads to any nontrivial changes in the vacuum expectation value of
the trace of the Energy - Momentum Tensor $<T_\mu^\mu>$.
To do this  it is
necessary to make explicit calculations in a covariant gauge
( or extract the divergences
with the help of some anzats) and put $<T_\mu^\mu> = 0$ that gives the
one-loop
conditions of Weyl invariance.
Even if these conditions are different from the standard effective equations,
it doesn't mean, of course, that the last are uncorrect.
In fact one has to check,
whether the given loop approximation in the theory with quantum metric
corresponds to the same approximation in the standard approach.
In other words one must
check that the $\alpha'$ is the parameter of the loop expansion even if the
two dimensional metric and sigma model coordinates are considered on
an equal footing. If this point is not satisfied whereas
the effective equations are different,
we arrive at the conditions of the Weyl invariance which are qualitatively
different from the standard ones but do not contradict them. Here we
show that it is just the case. It turns out that the one loop
conditions for the Weyl invariance of must be essentially modified if the
effects of quantum metric are taken into account. However the change of the
order of integrations disturbs  the standard
logical construction, related with the Weyl invariance conditions. If the
two dimensional metric is quantized then $\alpha'$ is no longer
the parameter of the loop expansion. Thus there is no small parameter
which can keep under control the contributions of higher loops, and
therefore in this case the "effective equations" have restricted sence.
For instance, one can construct such "effective equations" in $n$th
loop but these equations will not preserve the part corresponding to
the $n-1$ loop.
In spite of this it is believed that the study of the string theory
interacted with $2d$ gravity can be useful for the better understanding of
the general link between quantum gravity and the theory of strings.

If we deal with the one-loop approximation only, then the
difference between the divergences calculated in two different gauges
must be proportional to the classical equations of motion. Therefore one
can apply more comfortable conformal gauge to derive the
conditions of Weyl invariance and then verify the equivalence with
the covariant gauge. On this way we can compare the qualitative consideration
of Ref.'s \cite{31,32} with the result of direct calculation in the
covariant gauge.

     The action of closed boson string in the massless and tachyon
background fields has the form:
\beq
S = \int {d^2}\sigma \sqrt {g} \; \{ {1\over {2 \alpha'}}
{g^{\mu\nu}} G_{ij}(X)
\partial_{\mu}{X^i} \partial_{\nu}{X^j} + {1\over {\alpha'}}
{\varepsilon^{\mu\nu}\over {\sqrt {g}}} {A_{ij}}(X)
\partial_{\mu} {X^i} \partial_{\nu} {X^j}
+ B(X)R + T(X)\}                                          \label{1}
\eeq

     Here $i,j = 1,2,...,D; \;  {\mu, \nu} = 1,2; \;\;\;
  G_{ij}$  is the background
metric, $A_{ij}$ is the antisymmetric tensor background field, $B(X)$ is the
dilaton field, $T(X)$ is the tachyon background field. $R$ is the two
dimensional curvature, $X^{i}({\sigma})$ are the string coordinates.
At quantum level, since we do not follow the order of
integrations, the two dimensional metric is quantum field. That is why one can
regard (\ref{1}) as the action for the two dimensional quantum
gravity.

It is easy to see that two actions (\ref{1}) and (\ref{4x}) have
a very similar structure and hence  one can start directly with the
action (\ref{1}), regarding $g_{\mu\nu}$ as quantum field. In fact the
results (at the one-loop level) are qualitatively the same even if the
sum of (\ref{1}) and (\ref{4x}) is choosen as the starting action \cite{npsi}.
Let us notice that the theory with the action $S+S_{g}$ is equivalent
to the $D+1$ dimensional sigma model
\beq
S = \int {d^2}\sigma \sqrt {g} \;\{
 {1\over {2 \alpha'}} { g^{\mu\nu}} G_{ab}(Y)
\partial_{\mu}{Y^a} \partial_{\nu}{Y^b} +
{1\over{\alpha'}}{\varepsilon^{\mu\nu}\over {\sqrt {g}}}\; {A_{ab}}(Y)
\partial_{\mu} {Y^a} \partial_{\nu} {Y^b} + {\cal B}(Y)R + T(Y)\}  \label{3}
\eeq
with the special restrictions on the background fields
\cite{31,32,modphys,npsi}.
Here the indeces $a,b,..$ take the values $1,2,...,D+1$.
Below we deal only with the model (\ref{3}),
omitting the mentioned restrictions for the sake of brevity.

As far as we consider only the one-loop
renormalization, the gauge dependence of divergences have to be
proportional to the classical equations of motion.
For the sake of simplicity we write
only some (most relevant, because all the information is preserved)
combination of these equations, which follow from the action (\ref{3}).
$$
E_g=T - {\Box}{\cal B} = 0
$$
\beq
E_Y= - T + {\cal B}_{ab} {
g^{\mu\nu}} \partial_{\mu}{Y^a} \partial_{\nu}{Y^b} + {{\cal B}_a}
{H^a_{bc}} {\varepsilon^{\mu\nu}\over {\sqrt {g}}}
\partial_{\mu} {Y^b} \partial_{\nu} {Y^c} + {\alpha'}
{{\cal B}_a}{{\cal B}^a}R +
{\alpha'}{{\cal B}_a}{T^a} = 0                          \label{4}
\eeq
The indices $a,b,c$
near ${\cal B},\; T$ and $Y$ indicate to the covariant derivatives in
target $D + 1$ dimensional space. $H$ and $K$ (below) are the  torsion
and corvature tensors based on $G$ and $A$.

The renormalizability of the theory (\ref{3})
follows from the power counting consideration.
Indeed it is important that it is possible to preserve the diffeomorphism
invariance on the wourld sheet and also reparametrization invariance
in target space on quantum level. It is fairly
easy to see that we are able to
preserve both symmetries, because there
exists the covariant (i.e. dimensional)
regularization. It is also relevant that we can work in harmonic type
(covariant) gauge which is not sensitive to this regularization. However
for our purposes it is very convenient to apply the conformal gauge in
parallel.

If one introduces the conformal background gauge, then the action
(\ref{3})
can be expressed in the form of the $D+2$ dimensional nonlinear
sigma-model of the special form \cite{31,32}.
The two dimensional metric in this case is not quantized.
The conformal background gauge is introduced by the relation
\beq
 g_{\mu\nu} = e^{-2\rho}{\bar g}_{\mu\nu}                \label{5}
\eeq
where ${\bar g}_{\mu\nu}$ is some fixed background metric. In this
gauge the
action (\ref{3}) becomes
\beq
 S = \int {d^2}\sigma \sqrt {\bar g}\; \left\{
{1\over {2 \alpha'}} {{\bar g}^{\mu\nu}} G_{AB}(Z)
\partial_{\mu}{Z^A} \partial_{\nu}{Z^B} +
\frac{\varepsilon^{\mu\nu}}{\alpha'\sqrt{{\bar g}}}{A_{AB}}(Z)
\partial_{\mu}
{Z^A} {\partial}_{\nu} {Z^B} + {\cal B} R({\bar g}) + T \right\}
   \label{6}
\eeq
where $ A,B = 1,2,...,D + 2$ and
\beq
Z^{A} = ({\rho},\;{Y^a}),\;\;\;\;\;\;\;
G_{AB} = \left(\matrix{
              0                         &- 2{\alpha'}{\cal B}_b\cr
              - 2{\alpha'}{\cal B}_b    &{G_{ab}}
\cr}\right),\;\;\;\;\;\;\;
A_{AB} = \left(\matrix{
              0               &0\cr
              0               &{A_{ab}}
\cr}\right)
                                          \label{7}
\eeq
Below we shall use both (\ref{3}) and (\ref{6}) forms of the action.

The action (\ref{3}) is invariant under the general coordinate
transformations
in two dimensional and also in $D+1$ dimensional spaces.
However, since the two dimensional metric is quantized, it is natural to
expect that some extra symmetry takes place.
Such symmetry turns out to be the particular form of the
reparametrizations of the $D+2$ dimensional model (\ref{6}), which preserves
the block structure of the background metric (\ref{7}) \cite{anom}.
The generalised symmetry cause an arbitrariness of renormalization,
so the last is related to the reparametrizations and also to the
conformal transformation of the two-dimensional metric.

The calculation of
divergent part of the one loop effective action in the theory (\ref{3})
is performed in a way similar to the one
we have used above for the theory (\ref{4x}).
The difference is that here we apply
the background field method simultaneously in two  dimensional and
$D+1$  dimensional covariant form.

The background shift of the fields $g_{\mu\nu}, Y^a$ is
performed according to (\ref{11x}) and
\beq
Y^{a}\; {\rightarrow}\; {Y'^{a}} = {Y^{a}} +
{\sqrt{\alpha'}}\;{\pi}^{a}({\eta^{a}})            \label{14}
\eeq
The expansion (\ref{14}) supposes the standard
use of the normal coordinate method. The quantum field $\pi^a$ is
 parametrized
by the tangent vector $\eta^a$ in $D+1$ dimensional target space
(see \cite{npsi,anom} for the notations).

The more general form of the gauge fixing term, which  corresponds to
the covariant harmonic gauge is following:
\beq
 S_{gf} = - \int {d^2}\sigma \sqrt
{g}\;{\chi}_{\mu}\; {G^{\mu\nu}}\;{\chi}_{\nu} \label{16}
\eeq
where
${\chi}_{\mu}$ is the background gauge and $G^{\mu\nu}$ is the weight
operator.
\beq
{\chi}_{\mu} = {\nabla}_{\nu}{{\-h}_{\mu}^{\nu}} -
{\beta} {\nabla}_{\mu} h - {\gamma}_{a}{\nabla}_{\mu}Y^{a} -
E_{\mu}^{\rho\sigma} h_{\rho\sigma} - F_{{\mu}a}Y^{a},
\;\;\;\;\;\;\;
G^{\mu\nu} = {\tau} {g^{\mu\nu}} \label{17}
\eeq
Here ${\tau},\; {\beta},\; {{\gamma}_a}, \;E_{\mu}^{\rho\sigma},\;
F_{{\mu}a}$ are arbitrary functions (gauge parameters).
which  are taken in a
special way to provide the minimality of the total action.
\beq
 {\beta} = 0, \;\;\;\;\;
{\gamma}_{a} = -{\nu {\cal B}_a \over{{\cal B}}},\;\;\;\;\;
{\tau} = {{\cal B}\over{\nu}}                                 \label{18}
\eeq
There remains an arbitrariness
related with the parameter $\nu$ and also
the unessential one related with the functions
$E_{\mu}^{\rho\sigma}$ and
$F_{{\mu}a}$. The explicit calculation shows that the divergent part
of the one - loop effective action does not depend on these functions
just as in the pure gravity theory (\ref{4x}).

The bilinear part of the total action $S + S_{gf}$ has the form
\beq
(S + S_{gf})^{(2)} = \frac{1}{2}\int {d^2}\sigma \sqrt {g}\;({\bar
h}_{\mu\nu},\;h,\;\;\eta^a ) (\hat{H})
({\bar h}_{\alpha\beta},\;h,\;\;\eta^b)^T                   \label{19}
\eeq
where the self - adjoint operator ${\hat H}$ has the form (\ref{17x}).
Here ${\hat K}, {\hat L}^{\lambda},{\hat M}$
are c-number operators acting in the space of the fields
$({\bar h}_{\rho\sigma},\;h,\; {\eta}^{a})$.
The complete set of technical details are presented in \cite{npsi}.
Here we write down only the matrix ${\hat K}$.
\beq
{\hat K} = \left(\matrix{
\frac{{\cal B}}{2\nu} P^{\mu\nu,\;\alpha\beta}& 0 &0\cr
0 & 0 & -\frac{1}{2}{\cal B}_b\sqrt{\alpha'}\cr
0 & -\frac{1}{2}{\cal B}_a\sqrt{\alpha'} & - G_{ab} +
\frac{\nu\alpha'}{{\cal B}} {\cal B}_a {\cal B}_b \cr} \right)   \label{21}
\eeq

Here $P^{\mu\nu,\;\alpha\beta}$ is the projector to the traceless states.
It is important that the first term in ${\bar H}$ corresponds to
the propagator
of the theory (\ref{3}). Since ${\hat K}$ is inhomogeneous in $\alpha'$,
this constant does not play as the parameter of the loop expansion in the
theory with quantum metric.

The action of the Faddeev-Popov ghosts is defined in a usual way.
Summing up the contributions of $Tr \ln {\hat H}$ and the ghost part
we get \cite{modphys}
$$
{\Gamma}^{(1)}_{div} = - {1\over{\varepsilon}}
\int {d^2}\sigma \sqrt {g}\; \left\{ \left[
\;{{24 - D}\over 12} + {{\alpha'}\over{2}}
({\cal D}^{2}{\cal B}) - {{\alpha'}\over{2}}
{{\cal B}^{a}{\cal B}_{ab}{\cal B}^{b}
\over {{\cal B}^{c} {\cal B}_{c}}}\;\right]\;R\; -\;
 {{\nu}\over{{\cal B}}}({\Box}{{\cal B}} -T) +
\right.
$$
$$
\left.
+{1\over 2}
{{\cal K}^{a}_{a}} + {{\alpha'}\over{2}}{\cal D}^{2}T + {1 \over
{{\cal B}}^{c} {{\cal B}}_{c}}[\;{{\cal B}}^{a} T_{a} -
{{\alpha'}\over{2}} {\cal B}^{a} {T}_{ab}{\cal B}^{b}\;] - {1
\over {2 {\cal B}^{c} {\cal B}_{c}} } [ \;{1\over{\alpha'}}
{\Box}{\cal B} + {1\over{2}}{{\cal B}^b}{\Box{\cal B}_{b}} -
\right.
$$
\beq
\left.
-{1\over{2}} g^{\mu\nu}{{\cal B}}^{a}_{\mu} {{\cal B}}_{a \nu}
+ {{\cal B}}^{a} {{\cal B}}^{b}(g^{\mu\nu}H_{fbc} H^f_{\;\;ea}\;
{\partial}_{\mu}Y^{e} {\partial}_{\nu}Y^{c} + {{\cal K}}_{ab})-
2{{\cal B}}^{b} {{\cal B}}_{da}{\varepsilon^{\mu\nu}\over {\sqrt {g}}}
H^d_{\;\;eb}{\partial}_{\mu}Y^{a}{\partial}_{\nu}Y^{e}] \right\}  \label{29}
\eeq

One can compare the divergent part of the one - loop effective action
(\ref{29}) with the known results of other papers.
For example, the divergence of  Einstein type in (\ref{29}) has standard
form and does not depend on the gauge parameter
${\nu}$, that is just the expected result for the anomaly contribution.
Next, if one puts $D=0,\;\;\;{\cal B} = C_{1}{\Phi}$
and $H_{abc} = 0$ and so reduce the theory to the form (\ref{4x}), then
(\ref{29}) coincide with the expression, which was derived in \cite{29,ShB}
for pure gravity (\ref{4x}). Note also
that the ${\nu}$ - dependent terms in (\ref{29}) are proportional to the
equations of motion (\ref{4}), and hence disappear
on mass shell. One can consider this as some kind of control for the
correctness of the calculations.

Next feature of (\ref{29}) corresponds to the separation of quantum gravity
contributions, and it is not so obvious. In fact all the terms in
(\ref{29}), which are related with the contributions of quantum metric,
exhibit the dependence on $\nu$ or contain the factor of $({{\cal
B}_{a} {{\cal B}}^{a}})^{-1}$.  Thus removing all the ${\nu}$ -
dependent terms and also the terms which have ${{\cal B}}_{a} {{\cal
B}}^{a}$ in the denominators, we obtain the well-known result for the
$D+1$ dimensional sigma - model (\ref{3}) without quantum gravity \cite{3}.
Recently there were published the papers with calculations in the two
dimensional quantum gravity coupled to the linear sigma model
\cite{24b,EO}. The expression (\ref{29}) is in a good accord
with the results of these papers.

The expression (29) looks rather complicated and the appearance of
${{\cal B}}_{a}{{\cal B}}^{a}$ terms in some  denominators seems
strange, but in fact it is quite natural.
It turns out that this form of the divergences corresponds to the
relations between the geometric quantities like curvatures based on the
metric $G_{ab}$ and on the $D+2$ dimensional metric $G_{AB}$ (\ref{7}).
To understand this, we perform the parallel calculation in the
conformal gauge.

\vskip 8mm
\noindent
{\large \bf 6. Conformal gauge and the conditions of Weyl invariance}
\vskip 3mm

As it was already shown above, in the conformal gauge the theory (\ref{3})
is the $D+2$ dimensional nonlinear sigma model on the
background of purely classical two dimensional metric (\ref{6}).
We need the explicit expression for the one-loop divergences in
conformal gauge, because it gives the possibility to derive
the VEV of the Energy-Momentum Tensor in a more simple way.
In fact there is no need to make special  loop
calculations in this case. Since in the conformal gauge we are dealing
 with the ordinary
$D+2$ dimensional sigma model with the background fields $G$ and $A$
of special form (\ref{7}) it is possible to
use the well known result for the one-loop
divergences of the ordinary sigma-model. On this way we
 obtain these divergences
expressed in terms of the fields $T,\; {\cal B},\;
\;G_{AB}$ and $ A_{AB}\;$ (\ref{7}). Then the expression
for the divergences
can be rewritten in terms of $D+1$ geometrical quantities. The above
consideration essentially uses the universal form  of the divergences
of quantum field theory in an external field \cite{38}.

We have performed an independent calculation in conformal gauge and also
applied the standard result of Callan, Friedan, Martinec and Perry
\cite{call} together with the reduction formulas for the geometric
quantities \cite{npsi,anom}.
The results coincide and give the following
difference between two effective actions
(the symbol $c$ denotes
the use of the conformal gauge).
\beq
{\Gamma}^{(1c)}_{div}
- {\Gamma}^{(1)}_{div} = - {1\over{\varepsilon}}
\int {d^2}{\sigma} {\sqrt {g}}\; \{\;- \left[ { {\nu}\over{\cal B} } +
{1 \over {2 \alpha' {\cal B}^{a} {\cal B}_{a}}} \right] E_g\; + \;
{1 \over {2{\cal B}^{c} {\cal B}_{c}}}
\left[ {{\cal B}^{a}{\cal B}_{ab}{\cal B}^{b} \over {{\cal B}^{c} {\cal
B}_{c}}} -
({\cal D}^{2} {\cal B}) \right] E_Y\; \}                     \label{32}
\eeq

The expression for ${\Gamma}^{(1c)}_{div}$
is just the known divergent part of
the effective action for the ordinary (that is without quantum
gravity) sigma-model in the case of the restricted background
interaction fields (\ref{7}).  Thus we see that the denominators in
(\ref{29})
reflect the form of the curvature tensor in $D+2$ dimensional
target space with special form of the metric (\ref{7}) in this space.

The expressions for
 ${\Gamma}^{(1c)}_{div}$ and ${\Gamma}^{(1)}_{div}$ do not coincide,
nevertheless both are correct and moreover
verify the correctness of each other. To see this one can apply the
arguments of Ref. \cite{ShJ}. The general theorem \cite{39, VLT}
(see also \cite{38}) claims that the difference between two
effective actions derived
in two different gauges must vanishes on mass shell.
In particular, for the one-loop divergences this difference vanishes on the
classical equations of motion. Both
integrands of ${\Gamma}^{(1c)}_{div}$ and $ {\Gamma}^{(1)}_{div}$ are local
functionals of the same dimension as the equations of motion, and therefore
the difference
${\Gamma}^{(1c)}_{div} - {\Gamma}^{(1)}_{div}$ must be
the linear combination
of the equations of motion with the local coefficients, that is just
the case (\ref{32}). One can suppose that ${\Gamma}^{(1c)}_{div}$ and
${\Gamma}^{(1)}_{div}$ can be converted into each other after some
reparametrization transformation and that the corresponding
arbitrariness of the interaction fields have the well - known Killing
form.

At one loop all the divergences can be removed by the renormalization
of the background fields $G_{ab},\; A_{ab},\;{\cal B},\; T$ and hence the
renormalization of the two-dimensional metric is not necessary.
 From this point
of view the renormalization of the model under consideration is
qualitatively the same as for the ordinary sigma model (\ref{1}). However the
expressions for the beta functions are essentially
more complicated in our case.

Now we can formulate the conditions of Weyl invariance
at the one-loop level.
In the framework of the standard approach the theory can be formulated in
such a manner
that the Weyl invariance is the symmetry of the theory on both classical and
quantum levels. To do this, it is necessary to input the loop order
parameter $\alpha'$ into the starting action and thus divide this action
into two parts. One of these parts, that is classical action, is Weyl
invariant, and another one gives the nonzero contributions to the trace of
the Energy-Momentum Tensor $<T_{\mu}^{\mu}>$.
These contributions contains the additional $\alpha'$ factor. The one loop
corrections also give contributions of the same order, and therefore
it is possible to provide the Weyl invariance at the one loop level
if the corresponding conditions (effective equations) for the background
fields are introduced. This scheme can be extended to higher loops.
 In fact in a usual approach it is also supposed
the integration over the two-dimensional metric, but it is performed after
the integration over the sigma-model variables when the conditions of
Weyl invariance are already satisfied (see, for example, \cite{37}).
Then the integration over the metric is reduced to the summation over
topologies, that exactly corresponds to the anomaly free Weyl invariant
theory in two dimensions.

In the theory under discussion the two-dimensional geometry
possesses classical dynamics at classical level. However,  from general
point of view, if we want to develop a theory having relation to string,
the effective action has to be independent on the scale factor. Hence it
is necessary to formulate the conditions of the Weyl invariance on
quantum level in the theory with quantum gravity and thus check whether
there will be some real difference with the standard approach.

The standard way to derive $<T_\mu^\mu>$
is to start with the action (\ref{3}) and calculate the renormalization
of the composite operators \cite{tse,osb}.
However in our case we can apply more simple method.
The necessity to renormalize $g_{\mu\nu}$
does not appear in the model (\ref{3}) in both harmonic and
conformal gauges.  Hence, at one loop
level there is no big difference between two gauges
and one can apply
anyone of them to formulate the effective equations.
At the same time in the conformal gauge the model (\ref{3}) with quantum
gravity is equivalent to the ordinary sigma model in higher dimension
$D+2$ (\ref{6}). As far as we know the effective equations for
 ordinary sigma model and also the relations between the
geometrical quantities in dimensions $D+1$ and $D+2$,
it is not difficult to derive the effective equations for the
model (\ref{3}).

Since the introduction of the tachyon term leads to
the well known difficulty that is to the lack of equivalence between the
different approaches to the string theory (see for example \cite{37}),
we omit this term here.
Applying the reduction formulas \cite{npsi,anom}
we find the effective equations
for $G,\;A$ and $B$
background fields in the following form.
$$
 {\bar {\beta}}^G_{ab} = K_{ab} +
\frac{1}{{\cal B}^2}\;[\;{\cal B}_{ac}{\cal B}_b^c - {\cal B}^c{\cal
B}^d K_{acbd} + ({\cal D}^{2} {\cal B}) {\cal B}_{ab}]+
$$
$$
+\frac{1}{{\cal B}^4}[\frac{1}{4}({\cal B}^2)_{,a}({\cal B}^2)_{,b} -
\frac{1}{2}({\cal B}^2)_{,c}{\cal B}^c \;{\cal B}_{ab}\;]
- \frac{1}{4} H _a^{\;\;cd} H_{bcd} = 0
$$
$$
{\bar {\beta}}^A_{ab} = {\cal D}_c H^c_{\;\;ab} + \frac{1}{{\cal B}^2}
[2 {\cal B}^c {\cal B}_{d[a}H^d_{\;\;b]c} -
({\cal D}^2 \; {\cal B}\;{\cal B}^d +
{\cal B}^c{\cal B}^d\; {\cal D}_c) H_{dab}] +
+\frac{1}{{\cal B}^4}\; {\cal B}^c{\cal B}_{cd}{\cal B}^d {\cal
B}^e H_{eab} = 0
$$
$$
{1 \over {\alpha'}} {\bar {\beta}}^B
= {1 \over {\alpha'}} \frac{D-24}{48\pi^2}+
\frac{1}{16\pi^2}[4{\cal B}^2 - K -
\frac{1}{{\cal B}^2}({\cal B}_{ab}{\cal B}^{ab}
+{\cal B}^{a}{\cal B}^{b}K_{ab} + ({\cal D}^{2} {\cal B})^2)-
$$
\beq
- \frac{1}{{\cal B}^4}
( \frac{1}{4}G^{ab}{\cal B}^2)_{,a}({\cal B}^2)_{,b} - \frac{1}{4} {\cal B}^c
({\cal B}^2)_{,c} {\cal D}^{2} {\cal B}) + \frac{1}{12} H_{abc}H^{abc}] = 0
                                                                \label{33}
\eeq

The above equations are the conditions of the Weyl invariance in the
theory (\ref{3}) with the one-loop corrections.
So we observe that the effective equations in the
theory (\ref{3}) with quantum gravity are much more complicated and have
qualitatively different structure as compared with the ones that arise
in the framework of the standard approach. Since the arbitrariness of the
renormalization is restricted by the reparametrizations and
conformal transformation of the metric, there is
no any hope to reduce the equations
(\ref{33}) to standard form and therefore
the difference between two sets of effective equations is essential.
On the other hand, since we have used the harmonic gauge the starting model
(\ref{3}) is obviously renormalizable in it's original form and therefore
the difference between ours and standard sets of effective equations can not
be explained by the features of the noncovariant conformal gauge.
In this points our conclusions differ from the ones of Ref's \cite{31,32}.
Despite the difference with standard approach was already discussed above
let us tought this problem again.

The starting action of the theory with quantum gravity
is not invariant under the Weyl transformation, and the
two-dimensional metric in our theory has the nontrivial dynamics
already at the classical level. Since the two-dimensional metric
is quantized, the $\alpha'$ does not play the role of the loop expansion
parameter. Whereas the standard one-loop conditions of
Weyl invariance correspond to the first order in $\alpha'$, our
one-loop conditions of Weyl invariance correspond to the first order in
the expansion  (\ref{14}) which
essentially includes the metric part. Thus our one-loop approximation does
not correspond to the one-loop approximation in the standard approach.
In fact the difference can be seen already at the tree level, since we are
not able to extract the Weyl invariant classical part from the starting
action (\ref{3}) that is an important component of the standard approach.
At the same time the account of loop corrections leads to the effective
equations which strongly differ from the standard ones. In particular,
(\ref{33}) contains some higher derivative terms already at the one loop
level, whereas in the standard approach these terms appear only at second
loop.

The difference between our approach and the standard one
is that we do not follow the order of integrations. Since we perform the
integration over the metric simultaneously with the integration over the
sigma model coordinates, our loop expansion of the effective action
does not correspond to the expansion in the powers of $\alpha'$ and
that is why the resulting effective equations are different at any finite
loop. One can suppose that the difference between two approaches
disappear if we sum up the contributions of all the higher
loops and compare the full effective actions which arise in two approaches.
Unfortunately at the moment this is very far from the real possibilities.

\vskip 6mm
\noindent{\large \bf Conclusion}
\vskip 3mm

We have considered some aspects of the perturbative approach to the
two-dimensional quantum gravity. In the covariant gauge the dilaton
gravity (\ref{4x}) is quite similar to the dilaton models in $d=4$
(see, for example, \cite{spec}). The main difference is, of course,
the power counting renormalizability of gravity in $d=2$.
Moreover
in $d=2$ there is an additional identity for the quantum metric
(\ref{14x}) that leads to the special features of quantum corrections
like the strong dependence on the gauge fixing parameters. The use
of the covariant gauge quarantees the renormalizability of the theory.
That enables us to study the theory of string coupled to $2d$ gravity
and calculate an alternative effective equations for the background fields.

\vskip 6mm
\noindent{\large \bf Acknowledgments}

It is a pleasure for me to appreciate the collaboration and numerous
discussions with I.L.Buchbinder and S.D. Odintsov, who learned me
a lot of things related with the subject.
I am very grateful to M.Asorey, H.Kawai, Y.Kitazawa, Choonkyu Lee,
M.Ninomia, N.Sakai, A.A.Tseytlin and I.V.Tyutin for the useful discussions
and to A.G.Sibiryakov and A.T.Banin for collaboration.
The work
was supported in part by
Grant No RI1000 from The International Science Foundation and by
Russian Foundation for Fundamental Research, project no. 94-02-03234.


\end{document}